\documentclass[11pt]{article}

\usepackage{CJK}
\usepackage{color}
\usepackage{amsfonts}
\usepackage{mathrsfs}
\usepackage{amsmath,amssymb}

\begin{document}

\title{Information transfer in generalized probabilistic theories based on weak repeatability}

\vskip0.1in
\author{\small Zhaoqi Wu\thanks{Corresponding author. E-mail: wuzhaoqi\_conquer@163.com} $^{1,3}$, Shao-Ming Fei\thanks{Corresponding author. E-mail: feishm@cnu.edu.cn} $^{2,3}$,
Xianqing Li-Jost$^{3}$ and Lin Zhang$^{4}$\\
{\small\it  1. Department of Mathematics, Nanchang University, Nanchang 330031, P R China} \\
{\small\it  2. School of Mathematical Sciences, Capital Normal University, Beijing 100048, P R China}\\
{\small\it  3. Max-Planck-Institute for Mathematics in the Sciences, 04103 Leipzig, Germany}\\
{\small\it  4. Institute of Mathematics, Hangzhou Dianzi University, Hangzhou 310018, P R China}}
\date{}
\maketitle

{\bf Abstract} {\small } Information transfer in generalized probabilistic theories (GPT) is an important problem. We have dealt with the problem based on repeatability postulate, which generalizes Zurek's result to the GPT framework [Phys. Lett. A \textbf{379} (2015) 2694]. A natural question arises: can we deduce the information transfer result under weaker assumptions? In this paper, we generalize Zurek's result to the framework of GPT using weak repeatability postulate. We show that
if distinguishable information can be transferred from a physical system to a series of apparatuses under the weak repeatability postulate in GPT, then the initial states of the physical system must be completely distinguishable. Moreover, after each step of invertible transformation, the composite states of the composite system composed of the physical systems and the apparatuses must also be completely distinguishable.

\vskip 0.1 in



\vskip0.1in

\noindent {\bf 1. Introduction}

\vskip0.1in

In many interpretations of quantum mechanics, probabilities are treated as a primitive notion. For example, according to the Born rule, different outcomes accur according to a probability distribution related to the measurement operators \cite{NC}. During the past few years, quantum information theory ushered in a period of prosperity, but there is still many unsolved problems concerning the foundations of quantum mechanics. In this regard, many probabilistic frameworks beyond quantum theory have been put forward. Among them, a framework that is more general than quantum theory and classical theory from probabilistic perspective was introduced in \cite{BBLW1,BBLW2}, which is called generalized probabilistic theories (GPT), also termed convex operational theories in \cite{BA}, as in this framework the state space and the effect space are all assumed to be convex. Many physical quantities and information processing tasks such as distinguishability measures like fidelity, entropy, Tsirelson's bound, no-cloning, no-broadcasting, state discrimination, discord, steering, non-locality, coexistence of effects, compatibility of channels, incompatibility and simulability of observables, etc., are extensively reconsidered in GPT and it is found that many of them are possessed beyond quantum theory \cite{BBLW2}-\cite{AJ}.

In the aspect of quantum computation, the Born rule is very crucial to the computing power of a computing machine. In the duality quantum computer \cite{LGL1}-\cite{LGL2}, the power of duality quantum computer depends on the results of measurement on a partial wave function. Measurement postulates are also important in the foundations of quantum mechanics and some applications of quantum information \cite{SAS}-\cite{NK}.

The Born rule and state update (or `collapse') rule were employed to describe how quantum states change under measurement in current interpretations of quantum theory, while Schr\"odinger equation gurantees that each closed system will evolve under unitary transformations. However, unitary evolution postulate and wave-packet collapse postulate are in some sense irreconcilable. Many efforts have been made in trying to overcome this deficit. It was Zurek who deduced wave-packet collapse in the case of one apparatus based on unitary evolution and repeatability postulate \cite{ZUR1}, which provides a new perspective to tackle this problem in the foundations of quantum theory. Luo presented two new derivations by posing weak repeatability postulate or covariant condition for one apparatus case \cite{LUO}. The results in \cite{ZUR1} and \cite{LUO} have been extended from the setting of quantum theory to the framework of GPT in \cite{ZP} and \cite{WU1}, respectively. Furthermore, Zurek investigated the wave-packet collapse problem under a more general scenario where a series of apparatuses and the environment are taken into consideration, where repeatability postulate is imposed \cite{ZUREK}. The result in \cite{ZUREK} was extended to the framework of GPT in \cite{WU2}, and was reconsidered in \cite{WU3} in the framework of quantum theory by utilizing weak repeatability or covariant condition. In this paper, we generalize the result in \cite{ZUREK} to the framework of GPT by using weak repeatability postulate instead of repeatability postulate.

\vskip0.1in

\noindent {\bf 2. The framework of GPT}

\vskip0.1in

\noindent We first recall the operational framework of GPT (see \cite{BBLW2,GK3,ZP} for more details). For simplicity, we only consider finite GPT while the results also hold in general case by using some topological techniques.

(1) {\it State space.} The state space, denoted by $S$, is assumed to be a compact convex subset of a real, finite dimensional vector space and each $s\in S$ is called a state. The convexity assumption is imposed here in order to guarantee that a probabilistic mixture of two states is still a state.
The extreme points of $S$ are called pure states and denoted by $S_{pure}$.

(2) {\it Effect space.} We denote the set of affine functionals on $S$ by $\mathscr{A}(S)$, which is a partially ordered linear space, with the partial order defined by: $f\geq g$ iff $f(s)\geq g(s)$ for all $s\in S$. The order unit of $\mathscr{A}(S)$ is defined by an affine functional $\iota $ satisfying $\iota (s)=1$ for all $s\in S$. The unit interval $[0,\iota]$ is assumed to be the effect space $\mathscr{E}(S)$:
$$\mathscr{E}(S):=\{e\in \mathscr{A}(S)|0\leq e(s)\leq 1,\forall s\in S\}.$$
Note here that convex combinations of effects are again a valid effect. Moreover, there is a natural embedding of $S$ in $\mathscr{A}(S)^*$ (the dual space of $\mathscr{A}(S)$), given by $s\longmapsto \hat{s}$, where $\hat{s}(f)=f(s)$ for all $f\in \mathscr{A}(S)$. Hence, we can identify $s$ with $\hat{s}$, and regard $s(f)$ and $f(s)$ as the same.

(3) {\it Measurement.} A discrete measurement is assumed to be a set of effects $\{e_k\}_{k\in \mathscr{K}}$ satisfying $\sum_{k\in \mathscr{K}} e_k=\iota$, or equivalently, $\sum_{k\in \mathscr{K}} e_k(s)=1$ for all $s\in S$, which is denoted by $M=\{e_k\}_{k\in \mathscr{K}}$, where $\mathscr{K}$ is the outcome set. We shall denote the set of all measurements by $\mathcal{M}$. Note that $e_j(s)$ is the probability of getting $j$th outcome (say $a_j$, for instance) by a measurement $M$ in a state $s$. Note that we do not assume any concrete rule to give the outcome probability.
A discrete observable $\mathcal{O}$ is assumed to be a mapping from outcome set $\mathscr{K}$ to effect space $\mathscr{E}(S)$.

(4) {\it Transformation.} The physical transformations of a system are represented by a set of affine mappings $T:S\rightarrow S'$,
where $S$ and $S'$ are state spaces before and after the transformation, respectively. Note that not all such affine mappings are valid transformations in a particular theory.

(5) {\it State space for composite system.} Consider two systems $A$ and $B$, where $S_{A}$ and $S_{B}$ are the state spaces of the subsystems $A$ and $B$, respectively. To formulate the composite system $S_{AB}$, we impose the following assumptions:

(i) a joint state specifies a joint probability for each pair of effects $(e_A,e_B)$, where $e_A$ and $e_B$ are effects on system $A$ and $B$, respectively;

(ii) the joint probabilities obey the so-called no-signalling principle, i.e., the marginal probabilities for the outcomes of a measurement on $B$ does not depend on which measurement was conducted on $A$ and vice versa;

(iii) if the joint probabilities for all pairs of effects $(e_A,e_B)$ are specified, then the joint state is specified.

The minimal tensor product and maximal tensor product are defined as follows:
$$S_A\otimes_{min}S_B:=co\{s_A\otimes s_B: s_A\in S_A, s_B\in S_B\},$$
\begin{eqnarray*}S_A\otimes_{max}S_B&:=&\{\phi:\mathscr{E}(S_A)\times \mathscr{E}(S_B)\rightarrow \mathbb{R}: \phi\,\,\makebox{is a bilinear functional,}\,\,\phi(e_A,e_B)\geq 0,
\\&\,&\forall e_A\in \mathscr{E}(S_A)\,\, \makebox{and}\,\,e_B\in \mathscr{E}(S_B)\,\,\makebox{and}\,\,\phi(\iota_A,\iota_B)=1\},
\end{eqnarray*}
where $co$ denotes the convex hull, the product state $s_A\otimes s_B$ is defined by $(s_A\otimes s_B)(e,f)=s_A(e)s_B(f)$ for all pairs of effects $(e,f)\in \mathscr{E}(S_A)\times \mathscr{E}(S_B)$ and $\iota_A$ and $\iota_B$ are unit effects for systems $A$ and $B$, respectively.

The above-mentioned assumptions ensure that the composite system is a convex set which lies between the minimal tensor product and maximal tensor product, that is,
$$S_A\otimes_{min}S_B\subset S_A\otimes S_B\subset S_A\otimes_{max}S_B.$$

\noindent {\bf 3. Information transfer in GPT based on weak repeatability}

\vskip0.1in

In GPT, a closed physical system $S$ evolves under invertible transformations $\Gamma:S\rightarrow S$, just as a closed quantum system suffers unitary evolutions. By an invertible transformation, we mean a mapping $\Gamma'$ which satisfies that $\Gamma'(\Gamma(s))=\Gamma(\Gamma'(s))=s$ for all $s\in S$, where we denote such $\Gamma'$ by $\Gamma^{-1}$. This is a natural generalization of the concept of a unitary operator $U$ in quantum theory satisfying that $U^{\dag}U=UU^{\dag}=I$, where $I$ is the identity operator. And it is also reasonable to assume that any transformation on a physical system can be realized by an invertible transformation on a certain extended closed system \cite{ZP}. We adopt this assumption in our main result.

Recall that the fidelity between two states $s_{1}, s_{2}\in S$ in GPT is defined as \cite{GK3, ZP}:
\begin{equation}\label{eq1}
F(s_{1},s_{2})=\inf_{M }
F_{c}(p_{1}(M),p_{2}(M)),
\end{equation}
where inf is taken over all measurements $M=\{e_{i}\},$ $p_{1}(M)=\{e_{i}(s_{1})\}$ and $p_{2}(M)=\{e_{i}(s_{2})\},$
and $F_{c}(p,q)=\sum_{i}\sqrt{p_{i}q_{i}}$ is the classical fidelity between two probability distributions $p=\{p_i\}$
and $q=\{q_i\}$. Two states in GPT are said to be orthogonal if the fidelity between them is zero, which naturally
generalizes the concept of orthogonality for quantum states.

We now study information transfer problems in GPT based on weak repeatability postulate. Consider a physical system $S$ with an initial state $s^{u}$ and a series of measurement apparatuses $A_1,A_2,\cdots,A_k,\cdots$ whose initial states are prepared in $a_1,a_2,\cdots,a_k,\cdots$. Assume that the composite system $S\otimes A_1\otimes A_2\otimes \cdots \otimes A_k\otimes \cdots$ is so large such that it could be considered as a closed one. In other words, the following transformations $\Gamma_1, \Gamma_2, \cdots, \Gamma_k, \cdots$ imposed on the ``Physical system+Apparatuses" are all invertible ones.

We give our main result as follows:

{\bf Theorem 1}. If distinguishable information can be transferred from a physical system $S$ to a series of apparatuses $A_1, A_2,\cdots, A_k,\cdots$ under the weak repeatability postulate in GPT, then the initial states of the physical system $S$ must be completely distinguishable. Moreover, after each step of invertible transformation, the composite states of the composite system composed of the physical systems $S$ and the apparatus/apparatuses must also be completely distinguishable.

{\bf Proof.} First impose transformation $\Gamma_1$:
$$\label{eq1}s^{u}\otimes a_1 \otimes a_2 \cdots \otimes a_k \otimes \cdots \stackrel{\Gamma_1}{\longrightarrow}s_1^u \otimes
a_1^u \otimes a_2 \cdots \otimes a_k\otimes \cdots.$$

\noindent By the weak repeatability postulate, when applying $\Gamma_1$ to the outcome state $s_1^u$, the state of apparatus $A_1$ is still $a_1^u$ after the transformation:
$$s_1^u \otimes a_1 \otimes a_2 \otimes\cdots \otimes a_k \otimes \cdots\stackrel{\Gamma_1}{\longrightarrow}s_2^u \otimes a_1^u \otimes a_2\otimes\cdots\otimes a_k\otimes\cdots,$$
where $s_2^u$ is the outcome state of the physical system $S$ if $\Gamma_1$ is imposed again. Here is the fundamental difference between the weak repeatability postulate and the repeatability postulate: in the scenario of the former postulate, the system state will change when applying $\Gamma_1$ again (from $s_1^u$ to $s_2^u$, etc.), while in the scenario of the later one, the state of the physical system $S$ will not change after $\Gamma_1$ is implemented again (that is, $s_1^{u}\otimes a_1\otimes a_2\otimes\cdots a_k \otimes \cdots\stackrel{\Gamma_1}{\longrightarrow}s_1^u\otimes a_1^u\otimes a_2\cdots\otimes a_k\otimes \cdots$).

By imposing transformation $\Gamma_1$ repeatedly, we thus get
\begin{equation}\label{eq2}s_j^u\otimes a_1\otimes a_2\otimes\cdots\otimes a_k\otimes \cdots\stackrel{\Gamma_1}{\longrightarrow}s_{j+1}^u\otimes a_1^u\otimes a_2\cdots\otimes a_k\otimes \cdots, \quad j=0,1,2,\cdots,\end{equation}
where $s_0^u:=s^{u}$.

If the initial state of $S$ is $s_j^v$, similar arguments yield that
\begin{equation}\label{eq3}s_j^v\otimes a_1\otimes a_2\otimes\cdots\otimes a_k\otimes \cdots\stackrel{\Gamma_1}{\longrightarrow}s_{j+1}^v \otimes a_1^v\otimes a_2\otimes\cdots\otimes a_k\otimes \cdots, \quad j=0,1,2,\cdots,\end{equation}
where $s_0^v:=s^{v}$.

Recall that the fidelity in GPT admits the following properties \cite{GK3,ZP}:

(i) $F(s_1\otimes t_1, s_2\otimes t_2)\leq F(s_1,s_2)F(t_1,t_2)$, for all $s_1, s_2\in S_A$ and $t_1, t_2\in S_B$;

(ii) $F(s_1, s_2)=F(s_1\otimes t, s_2\otimes t)$, for all $s_1, s_2\in S_A$ and $t\in S_B$;

(iii) $F(\Gamma(s_{1}), \Gamma(s_{2}))=F(s_{1},s_{2})$, and $s_1,s_2\in S_A$,

\noindent where $S_{A}$ and $S_{B}$ are two state spaces and $\Gamma$ is any invertible transformation in GPT.

Since $\Gamma_1$ is invertible, combining Eqs. (\ref{eq2}), (\ref{eq3}) and using the properties (i)-(iii), we obtain
\begin{eqnarray}\label{eq4}
F(s_j^u,s_j^v)&=&F(s_j^u\otimes a_1\otimes a_2\otimes\cdots\otimes a_k\otimes \cdots,s_j^v\otimes a_1\otimes a_2\otimes\cdots\otimes a_k\otimes \cdots)  \nonumber \\
&=& F(s_{j+1}^u\otimes a_1^u\otimes a_2\otimes\cdots\otimes a_k\otimes \cdots, s_{j+1}^v\otimes a_1^v\otimes a_2\otimes\cdots\otimes a_k\otimes \cdots)
\nonumber  \\
&=& F(s_{j+1}^u\otimes a_1^u,s_{j+1}^v\otimes a_1^v)  \nonumber \\
&\leq & F(s_{j+1}^u, s_{j+1}^v)F( a_1^u, a_1^v), \quad j=0,1,2,\cdots.\label{eq5}
\end{eqnarray}
Noting that $s_0^u=s^{u}$ and $s_0^v=s^{v}$, it follows from iteration of Eq.(\ref{eq4}) that
\begin{equation}\label{eq5}F(s^u,s^v)\leq  F(s_j^u, s_j^v)F^j(a_1^u, a_1^v), \quad j=0,1,2,\cdots.\end{equation}

It is reasonable to assume that the states $a_1^u$ and $a_1^v$ are different, otherwise no information will be obtained from the measurement. Hence we have
$F(a_1^u, a_1^v)<1$. Since $F(s_j^u, s_j^v)\leq 1$, by letting $j\rightarrow \infty$ in Eq. (\ref{eq5}), we obtain
$F(s^u,s^v)=0$. This shows that if we allow $\Gamma_1$ repeat infinite times, $s^u$ and $s^v$ must be completely distinguishable.

We proceed to consider a second transformation $\Gamma_2$:
$$s_1^u\otimes a_1^u\otimes a_2\otimes\cdots\otimes a_k\otimes \cdots\stackrel{\Gamma_2}{\longrightarrow}\tilde{s}_1^u\otimes a_1^u\otimes a_2^u\otimes\cdots\otimes a_k\otimes \cdots.$$
By the weak repeatability assumption, the composite state will suffer the following change when $\Gamma_2$ is implemented again:
$$\tilde{s}_1^u\otimes a_1^u\otimes a_2\otimes\cdots\otimes a_k \cdots\stackrel{\Gamma_2}{\longrightarrow}\tilde{s}_2^u\otimes a_1^u\otimes a_2^u\otimes\cdots\otimes a_k\otimes \cdots.$$

Therefore, repeated implementation of $\Gamma_2$ leads to
\begin{equation}\label{eq6}\tilde{s}_j^u\otimes a_1^u\otimes a_2\otimes\cdots\otimes a_k\otimes\cdots\stackrel{\Gamma_2}{\longrightarrow}\tilde{s}_{j+1}^u\otimes a_1^u\otimes a_2^u\otimes \cdots\otimes a_k\otimes \cdots,\quad j=0,1,2,\cdots,\end{equation}
where $\tilde{s}_0^u:=s_1^u$. And for the $``v"$ counterpart, we just obtain
\begin{equation}\label{eq7}\tilde{s}_j^v\otimes a_1^v\otimes a_2\otimes \cdots\otimes a_k\otimes\cdots \stackrel{\Gamma_2}{\longrightarrow}\tilde{s}_{j+1}^v\otimes a_1^v\otimes a_2^v\otimes\cdots a_k\otimes\cdots ,\quad
j=0,1,2,\cdots,\end{equation}
and $\tilde{s}_0^v:=s_1^v$.

Applying the properties (i)-(iii), it follows from (\ref{eq6}) and (\ref{eq7}) that
\begin{eqnarray}\label{eq8}
F(\tilde{s}_j^u\otimes a_1^u,\tilde{s}_j^v\otimes a_1^v)&=&F(\tilde{s}_j^u\otimes a_1^u\otimes a_2\otimes\cdots\otimes a_k\otimes\cdots,\tilde{s}_j^v\otimes a_1^v\otimes a_2\otimes \cdots\otimes a_k\otimes\cdots)
\nonumber \\
&=& F(\tilde{s}_{j+1}^u\otimes a_1^u\otimes a_2^u\otimes \cdots\otimes a_k\otimes \cdots, \tilde{s}_{j+1}^v\otimes a_1^v\otimes a_2^v\otimes\cdots a_k\otimes\cdots)
\nonumber  \\
&=& F(\tilde{s}_{j+1}^u\otimes a_1^u\otimes a_2^u, \tilde{s}_{j+1}^v\otimes a_1^v\otimes a_2^v)
\nonumber \\
&\leq & F(\tilde{s}_{j+1}^u\otimes a_1^u, \tilde{s}_{j+1}^v\otimes a_1^v)F(a_2^u, a_2^v), \quad j=0,1,2,\cdots,\label{eq9}
\end{eqnarray}
which implies that
\begin{equation}\label{eq9}F(s_1^u\otimes a_1^u,s_1^v\otimes a_1^v)\leq F(\tilde{s}_{j}^u\otimes a_1^u, \tilde{s}_{j}^v\otimes a_1^v)F^j(a_2^u,a_2^v), \quad j=0,1,2,\cdots.\end{equation}

Similarly, $F(a_2^u, a_2^v)<1$ since distinguishable information from apparatus $A_2$ should be extracted. Letting $j\rightarrow \infty$ in Eq. (\ref{eq9}), we obtain $F(s_1^u\otimes a_1^u,s_1^v\otimes a_1^v)=0$.
This shows that if we allow $\Gamma_2$ be conducted repeatedly, $s_1^u\otimes a_1^u$ and $s_1^v\otimes a_1^v$ must be completely distinguishable.

We can continue this process and implement other transformations $\Gamma_3,\cdots,\Gamma_k,\cdots$ and derive corresponding results. In summary, we obtain that
\begin{equation}\label{eq10}F(s^u,s^v)=F(s_1^u\otimes a_1^u,s_1^v\otimes a_1^v)=F(\tilde{s}_1^u\otimes a_1^u\otimes a_2^u,\tilde{s}_1^v\otimes a_1^v\otimes a_2^v)=\cdots=0,
\end{equation}
which completes the proof.
\rightline{$\square$}

{\bf Remark 1}. In the derivations based on repeatability postulate in \cite{WU2}, it is assumed that same distinguishable records are left on each measurement apparatus, i.e., $F(a^u_1,a^v_1)=\cdots =F(a^u_k,a^v_k)=\cdots<1$. In the proof of Theorem 1, we only need $F(a^u_i,a^v_i)<1$, $i=1,2,\cdots,k,\cdots$, while the fidelities need not be the same.

It is worth pointing out here that the fidelity in GPT only satisfies sub-multiplicity in general, so we can only obtain Eq. (\ref{eq10}). Nevertheless, if one restricts the GPT to quantum theory, it is well-known that quantum fidelity admits multiplicity, that is, $F(\rho_1\otimes \sigma_1,\rho_1\otimes \sigma_2)=F(\rho_1,\sigma_1)F(\rho_2,\sigma_2)$ for all quantum states $\rho_1,\rho_2\in D(H_1)$ and $\sigma_1,\sigma_2\in D(H_2)$, where $D(H_1)$ and $D(H_2)$ are spaces of all density operators on Hilbert spaces $H_1$ and $H_2$, respectively. In this case, by the arguments in Theorem 1, we can get the following:
\begin{equation}\label{eq11}F(\rho^u,\rho^v)=F(\rho_1^u,\rho_1^v)=F(\tilde{\rho}_1^u,\tilde{\rho}_1^v)=\cdots=0.
\end{equation}

In other words, if distinguishable information can be transferred from a quantum system $S$ to a series of apparatuses $A_1, A_2,\cdots, A_k,\cdots$ under the weak repeatability assumption in quantum theory, then the initial states of the quantum system $S$ must be completely distinguishable. Moreover, the states of the quantum system $S$ in each step of the unitary evolution must be completely distinguishable, which is similar to the case in \cite{WU3}.

Unlike weak repeatability assumption, if one adopts covariant condition to consider the information transfer problem under the scenario of a series of apparatuses in GPT, one can not derive Eq. (10) but only $F(s^u,s^v)=0$, which makes no difference when only one apparatus is considered, although in quantum case, Eq. (11) holds if covariant condition is assumed \cite{WU3}.

\vskip0.1in

\noindent {\bf 4. Conclusions}

\vskip0.1in

The framework of generalized probabilistic theories (GPT) is a very powerful tool in studying the foundations of quantum physics. Many recent results have improved our understanding of the informational properties of quantum theory. By adopting weak repeatability postulate, we have provided an analytical derivation of information transfer in generalized probabilistic theories.

Wave-packet collapse is a mysterious phenomenon which plays an important role in different interpretations of quantum theory. Recently, the authors in \cite{SCM} derived a unified probability rule that subsumes both the Born rule and the collapse rule, and showed that this more fundamental probability rule can provide a rigorous foundation for informational interpretations of quantum theory. It is worth investigating the foundational problems and the information properties of quantum theory from the perspective of scrutinizing them in a broad framework. Our results may highlight further investigations on GPT framework and information transfer in quantum theory.

\subsubsection*{Acknowledgements}
Zhaoqi Wu would like to express his sincere gratitude to Prof. Shunlong Luo for fruitful discussions.
This work was supported by National Natural Science Foundation of China (Grant Nos. 11701259, 11461045, 11675113, 11301124), the China Scholarship Council (Grant No.201806825038), the Key Project of Beijing Municipal Commission of Education under No. KZ201810028042, Natural Science Foundation of Zhejiang Province of China (LY17A010027), and the cross-disciplinary innovation team building project of Hangzhou Dianzi University. This work was completed while Zhaoqi Wu was visiting Max-Planck-Institute for Mathematics in the Sciences in Germany.


\end{document}